\ifx\pdfoutput\undefined\else\pdfoutput=1\fi
\documentclass[aip,amsmath,amssymb,graphicx,reprint]{revtex4-1}

\usepackage{graphicx}
\usepackage{float}

\begin{document}

\title{Mapping Source-Resolved Phase-Noise Transfer in Soliton Microcombs}

\author{Zhongan Zhao}
\email{zhzhao@dtu.dk}
\affiliation{Department of Electrical and Photonics Engineering, Technical University of Denmark (DTU), DK-2800 Kgs. Lyngby, Denmark}

\author{Aleksandr Razumov}
\affiliation{Department of Electrical and Photonics Engineering, Technical University of Denmark (DTU), DK-2800 Kgs. Lyngby, Denmark}

\author{Viswanathan Sankar}
\affiliation{Department of Electrical and Photonics Engineering, Technical University of Denmark (DTU), DK-2800 Kgs. Lyngby, Denmark}

\author{Jasper Riebesehl}
\affiliation{Department of Electrical and Photonics Engineering, Technical University of Denmark (DTU), DK-2800 Kgs. Lyngby, Denmark}

\author{Darko Zibar}
\affiliation{Department of Electrical and Photonics Engineering, Technical University of Denmark (DTU), DK-2800 Kgs. Lyngby, Denmark}

\date{\today}

\begin{abstract}
Phase noise limits the coherence and stability of soliton microcombs, yet its origin is difficult to trace because multiple noise sources act simultaneously. It is often represented by common-mode and repetition-rate components, but how each physical source contributes to these components remains unclear. We combine subspace tracking with multi-source Ikeda-map simulations, switching each source and the Raman nonlinearity on and off to isolate its contribution. For a spectrally symmetric soliton, pump phase noise projects entirely onto common mode, whereas shot noise and amplified spontaneous emission (ASE) drive repetition-rate noise. In the Raman-dominated shifted state, pump phase noise is coherently transferred from common mode to repetition-rate noise, without the Raman response adding fluctuations of its own. Across the added-source configurations, Raman-mediated transfer of pump phase noise remains the dominant repetition-rate contribution. Shot noise, ASE, relative intensity noise (RIN), and thermorefractive noise (TRN) make only small additional contributions to the repetition-rate PSD while raising the common-mode floor. No additional growing component is resolved in any configuration examined, indicating that the two-component description remains sufficient. The intracavity dynamics thus do not merely carry noise but actively partition it, providing a mechanistic basis for low-noise microcomb design.
\end{abstract}


\maketitle

\section{Introduction}

Optical frequency combs (OFCs) consist of mutually coherent and equally spaced spectral lines, providing a phase-coherent link between the radio-frequency (RF) and optical domains\cite{diddams2020optical}. Microresonator frequency combs (microcombs) are particularly attractive because they combine high repetition rates, low power consumption, and compatibility with photonic integration. Under continuous-wave (CW) pumping, dissipative Kerr solitons (DKSs), sustained by a double balance---between anomalous dispersion and Kerr nonlinearity, and between cavity loss and external driving---enable stable and coherent microcomb generation\cite{Dissipative}.

The phase-coherent link provided by OFCs underpins precision metrology\cite{udem2002optical} and high-resolution spectroscopy\cite{picque2019frequencyspectroscopy}. At the system level, Kerr microcombs have enabled massively parallel coherent laser ranging (LiDAR)\cite{trocha2018ultrafastLIDAR,riemensberger2020massivelyLIDAR}, terabit- and petabit-scale coherent optical transmission\cite{marinpalomo2017microresonatorCOMM,jorgensen2022petabitCOMM}, and comb-driven data-center interconnects\cite{rizzo2023massivelyDCI,shu2022microcombDrivenSiP}. Across these applications, performance depends on both the phase stability of individual comb lines and the stability of their relative spacing. Phase noise and linewidth quantify the practical coherence of optical oscillators\cite{schawlow1958infrared,henry2003theory}, while the phase-noise power spectral density (PSD) resolves the spectral distribution of phase fluctuations and is therefore essential for characterizing and engineering low-noise optical sources\cite{di2010simple}.

High-resolution phase noise and linewidth characterization commonly relies on heterodyne techniques, which down-convert optical phase fluctuations to the radio-frequency domain. This principle is employed by dual-laser heterodyne\cite{kessler2012duallaser1,lee2013spiralduallaser2,pavlov2018narrowdual3}, delayed self-heterodyne\cite{okoshi1980novelDSHI1,riebesehl2025interferenceDSHI2,zhao2022narrowDSHI3}, and self-homodyne schemes\cite{ludvigsen1998laserhomo,gundavarapu2019subhomo2,chauhan2021visiblehomo3}. For frequency combs, such methods are usually applied after spectrally selecting an individual comb line, which is then treated as a single-wavelength laser to obtain its linewidth or phase-noise PSD\cite{kim2016ultralowkim}. This provides a useful benchmark for the coherence of selected lines. However, describing each line in isolation discards the inter-line correlations that carry the global structure of the comb noise: neighboring lines share much of their phase fluctuation, and no single line reveals whether fluctuations are common to all lines or differ between them. This is precisely the distinction between common-mode and repetition-rate noise.

The global phase noise of a frequency comb is commonly described in terms of common-mode and repetition-rate fluctuations, captured by the elastic-tape model\cite{telle2002kerrelastic}. The common-mode component represents phase fluctuations shared by all comb lines, while the repetition-rate component represents fluctuations of the line spacing whose contribution grows with comb-line index. Several schemes have been developed to extract these two components independently\cite{paschotta2006opticalCEOnoise,helbing2003carrierCEOnoise2,wildi2023sidebandrf1,shirpurkar2024opticalrf2}, providing a compact description of global comb phase-noise dynamics. The field has since progressed toward joint multi-line measurements, where phase-correlation matrices quantify the inter-line correlation structure across the comb spectrum\cite{brajato2019opticaldarkovictor}. More recently, subspace tracking has been applied to multi-line phase-noise data to separate and quantify the dominant phase-noise components\cite{razumov2023subspace}. Demonstrations in electro-optic (EO) combs, nonlinearly broadened EO combs, and mode-locked lasers have established its applicability across distinct comb platforms\cite{holgerEO,razumov2026nonlinearBroadeningEO,alekModelock}.

Soliton microcombs, however, exhibit noise-coupling pathways that differ fundamentally from those of mode-locked lasers and EO combs. The latter are tied to gain dynamics and RF modulation, respectively, whereas soliton microcombs arise from Kerr-parametric four-wave mixing in a passive resonator driven by a single-frequency CW pump. Pump phase fluctuations can therefore perturb the soliton state directly\cite{del2011octaveCWdrive}. Soliton microcombs have been realized in resonator platforms based on silicon nitride, silica, lithium niobate, and other crystalline materials\cite{karpov2016raman,yi2017single,he2019selfLiNbO3,herr2014temporalDKS}. The dispersion and nonlinear responses of these platforms depend on both material properties and device geometry. Higher-order nonlinear effects, notably the Raman self-frequency shift and dispersive-wave recoil, further modify the soliton group velocity and add timing jitter to the repetition rate\cite{yang2016spatialREDRAMAN,karpov2016raman}. Thermodynamic temperature fluctuations in the resonator mode volume shift the cavity resonance through the thermo-optic effect, giving thermorefractive noise (TRN), which broadens the comb-mode linewidths\cite{gorodetsky2004fluctuations,huang2019thermorefractive,drake2020thermal}. With the pump frequency fixed, TRN acts on the soliton as a fluctuating pump--cavity detuning; thermal-expansion and thermomechanical channels are weaker and are not included here. Pump phase noise, shot noise, amplified spontaneous emission (ASE) from an erbium-doped fiber amplifier (EDFA), relative intensity noise (RIN), and TRN enter the modeled intracavity dynamics through distinct pathways. Disentangling their effects is therefore necessary to trace the common-mode and repetition-rate noise to their physical origins.

Prior work showed that the repetition rate becomes first-order insensitive to pump–cavity detuning through the combined Raman self-frequency shift and dispersive-wave recoil\cite{yi2017single}, quantified the resulting dispersive-wave noise limits on the repetition rate\cite{yang2021dispersive}, and characterized how the Raman response, shot noise, and RIN shape the comb-line linewidth distribution and its red-shifted quiet point\cite{lei2022optical}. The source-to-component mapping therefore remains lacking: which components dominate the multi-line phase dynamics, which physical sources drive them, and whether the elastic-tape model remains sufficient under nonlinear cavity dynamics. A cavity nonlinearity that injects no fluctuations of its own can still move noise from one component into another. In an operating microcomb, however, these sources act simultaneously, and their individual projections onto common-mode and repetition-rate noise cannot be separated.

Here we resolve how the phase noise of a soliton microcomb arises. Using a controllable multi-source Ikeda-map framework together with subspace tracking, we construct a source-resolved map of phase-noise transfer that follows each perturbation from its physical origin to the common-mode and repetition-rate components of the comb. The source set spans optical, quantum, and thermodynamic origins, with TRN modeled as a colored process whose variance and correlation time follow from the resonator material and geometry. By switching individual noise sources and the Raman nonlinearity on and off, an intervention that is not accessible experimentally, we isolate the pathway of each contribution. This reveals which perturbation feeds each component and makes the counterintuitive routing concrete: the Raman self-frequency shift, though it adds no noise of its own, redistributes pump fluctuations from the common mode into the repetition rate. This attribution identifies the dominant coherence-limiting contribution in each configuration and the source or intracavity pathway that should be mitigated first. The resulting picture makes the intracavity dynamics themselves a design variable: they partition fluctuations among the comb's phase-noise components, so low-noise microcomb design can target the transfer pathway as well as the sources that feed it.

\section{Theory}
\subsection{Ikeda-map model}

We model the intracavity field evolution using an Ikeda-map framework that separates each cavity roundtrip into continuous propagation within the cavity and a discrete boundary update at the coupling interface. During the $m$-th roundtrip, the intracavity envelope $E_m(z,t)$ propagates according to the generalized nonlinear Schr\"{o}dinger equation (GNLSE)\cite{AGRAWAL201927Raman}:

\begin{equation}
\label{eq:propagation}
\begin{aligned}
\frac{\partial E_m(z,t)}{\partial z} = \left( -\frac{\alpha}{2} - i\frac{\beta_2}{2} \frac{\partial^2}{\partial t^2} \right) E_m  \\+ i\gamma \left[ (1-f_R)\,|E_m|^2 + f_R\,|E_m|^2_{R} \right] E_m,
\end{aligned}
\end{equation}

where $\alpha$ is the propagation loss, $\beta_2$ is the group-velocity dispersion coefficient, $\gamma$ is the Kerr nonlinear coefficient, and $f_R$ is the fractional Raman contribution to the total nonlinearity. The first term in square brackets is the instantaneous Kerr response. The second term, $|E_m|^2_{R} = \int_0^{\infty} h_R(t')\,|E_m(z,t-t')|^2\,dt'$, is the delayed Raman-weighted intensity, with the response function $h_R(t)$ defined in the Numerical implementation section. Setting $f_R = 0$ recovers the Kerr-only GNLSE, enabling direct comparison between simulations with and without the Raman response. Equation~\eqref{eq:propagation} is solved numerically via the split-step Fourier method.

After propagation through the cavity of length $L$, the field is recombined with the noisy pump at the coupling interface. The boundary condition for the $(m+1)$-th roundtrip reads:

\begin{equation}
\label{eq:boundary}
\begin{aligned}
E_{m+1}(0,t) = \sqrt{\theta} \, E_{\mathrm{in},m}(t) + \sqrt{1-\theta} \, E_m(L,t) \, e^{-i(\delta_0 +\, \delta_m^{\mathrm{TRN}})} \\+ \delta E_m^{\mathrm{(shot)}}(t),
\end{aligned}
\end{equation}

where $\theta$ is the power coupling coefficient and $\delta_0$ is the roundtrip phase detuning, to which $\delta_m^{\mathrm{TRN}}$ adds the thermorefractive fluctuation of the cavity resonance; its statistics are given in the Numerical implementation section. The shot-noise term represents zero-mean complex vacuum fluctuations entering through the cavity dissipation channels and sets the quantum-noise floor of the system. Its discrete normalization is specified in the Numerical implementation section. The noisy launched pump field $E_{\mathrm{in},m}(t)$, incorporating phase noise, intensity noise, and ASE, takes the form:

\begin{equation}
\label{eq:input_field}
E_{\mathrm{in},m}(t) = \sqrt{\bar{P}_{\mathrm{in}} (1 + \delta\eta_m)} \exp(i\phi_m) + \delta E_m^{\mathrm{(ASE)}}(t),
\end{equation}

where $\bar{P}_{\mathrm{in}}$ is the mean pump power incident on the resonator. The accumulated phase $\phi_m$ represents pump phase noise associated with the Lorentzian linewidth $\Delta\nu$\cite{mandel1996opticalWiener}. The term $\delta\eta_m$ is the dimensionless relative-intensity fluctuation specified by the pump RIN spectrum $S_{\mathrm{RIN}}(f)$. The additive field $\delta E_m^{\mathrm{(ASE)}}(t)$ represents EDFA ASE, with its level determined by the amplifier gain and noise figure\cite{agrawal2012fiberEDFA}. The corresponding discrete sampling rules for all noise sources are summarized in the Numerical implementation section.

\begin{figure*}[!tbp]
    \centering
    \includegraphics[width=\textwidth]{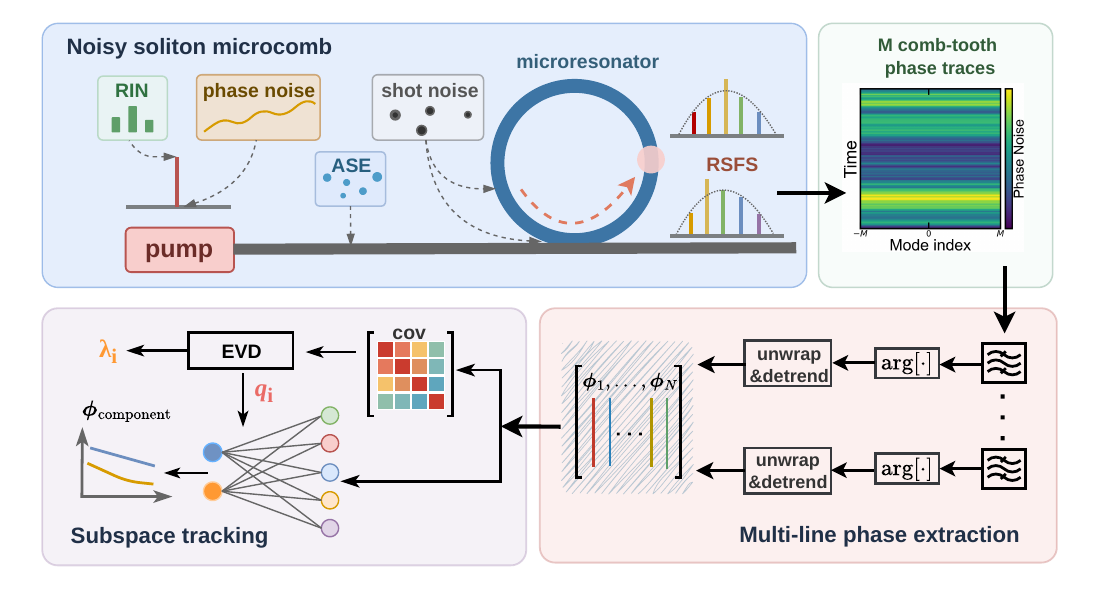}
    \caption{Simulation and analysis workflow for source-resolved phase-noise mapping. RSFS, Raman self-frequency shift; PN, phase noise; PSD, power spectral density; EDFA, erbium-doped fiber amplifier; ASE, amplified spontaneous emission; RIN, relative intensity noise.}
    \label{fig:pipeline}
\end{figure*}

\subsection{Elastic-tape model and subspace tracking}

The elastic-tape model serves as the reference description of comb-line phase noise\cite{telle2002kerrelastic}. We use the generalized form
\begin{equation}
\label{eq:elastic_tape}
\phi_n(t) = \phi_{\mathrm{cm}}(t) + n\,\phi_{\mathrm{rep}}(t) + \phi_{\mathrm{res}}(n,t),
\end{equation}
where $n$ is the comb-line index relative to the pump, $\phi_{\mathrm{cm}}(t)$ and $\phi_{\mathrm{rep}}(t)$ are the common-mode and repetition-rate phase-noise terms, and $\phi_{\mathrm{res}}(n,t)$ collects components outside their flat and linear line-index profiles. The standard elastic-tape model is recovered for $\phi_{\mathrm{res}}=0$. Retaining the residual term allows subspace tracking to identify significant phase-noise structure outside that basis.

For a selected set of $M$ comb lines with physical indices $\{n_1,\ldots,n_M\}$ defined relative to the pump, we collect the corresponding phase traces as $\boldsymbol{\phi}_{\mathrm{lines}}(t)=[\phi_{n_1}(t),\phi_{n_2}(t),\dots,\phi_{n_M}(t)]^{T}$ and write a low-dimensional component representation,
\begin{equation}
\label{eq:linear_model}
\boldsymbol{\phi}_{\mathrm{lines}}(t) = \mathbf{H}\boldsymbol{\phi}_{\mathrm{comp}}(t),
\end{equation}
where $\boldsymbol{\phi}_{\mathrm{comp}}(t)$ is a $P$-dimensional vector of effective phase-noise components, and $\mathbf{H}$ is an $M\times P$ coefficient matrix that maps these components to the selected comb lines. Each column of $\mathbf{H}$ describes the line-index profile of one component. The first two columns correspond to the flat common-mode and linear repetition-rate profiles, while additional columns describe residual line-index profiles.

After detrending and mean-centering the selected traces, denoted by $\tilde{\boldsymbol{\phi}}_{\mathrm{lines}}(t_k)$, we form the sample covariance matrix,
\begin{equation}
\label{eq:covariance}
\mathbf{C}_{\phi} = \frac{1}{K-1}\sum_{k=1}^{K}
\tilde{\boldsymbol{\phi}}_{\mathrm{lines}}(t_k)\tilde{\boldsymbol{\phi}}_{\mathrm{lines}}^{T}(t_k),
\qquad
\mathbf{C}_{\phi} = \mathbf{Q}\mathbf{\Lambda}\mathbf{Q}^{T},
\end{equation}
where $K$ is the number of time samples, the columns of $\mathbf{Q}$ are the eigenvectors, and $\mathbf{\Lambda}$ contains the corresponding eigenvalues. Each eigenvector gives one component's profile over the selected traces, while its eigenvalue gives the corresponding phase variance. Following Ref.~\cite{razumov2023subspace}, we set $\hat{\mathbf{H}}=\mathbf{Q}_{P}$, classify the retained eigenvectors as common-mode, repetition-rate, or residual components according to their line-index profiles, and obtain the component traces as $\hat{\boldsymbol{\phi}}_{\mathrm{comp}}(t_k)=\mathbf{Q}_{P}^{T}\tilde{\boldsymbol{\phi}}_{\mathrm{lines}}(t_k)$. Their PSDs and variances quantify the spectral content and relative weight of each component.

Because the eigendecomposition does not encode the absolute comb-line indices, the eigenvector entries must be mapped to the known indices $\{n_i\}$ to recover the physical phase-noise profiles. When one eigenvector is flat, orthogonality places the zero crossing of a linear eigenvector at the mean selected index. For each simulated case, the window center and span are chosen together to sample the corresponding line-resolved noise profile approximately symmetrically about its minimum. This case-specific choice simplifies the physical reference but is not required; an off-center window can be re-referenced using the known indices. The number of selected comb lines should exceed the anticipated number of phase-noise components. Using more lines provides additional sampling of the component profiles across the comb.

\section{Numerical results}

Using controlled Ikeda-map simulations, we determine how individual physical noise sources project onto the resolved phase-noise components and assess whether subspace tracking can recover the resulting component structure. Independent source loading isolates the source-to-component pathways examined below.

\begin{figure}
    \centering
    \includegraphics[width=1\linewidth]{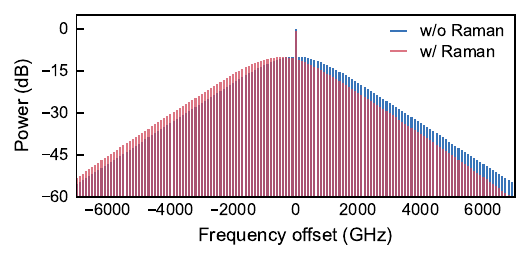}
    \caption{Steady-state soliton spectra without and with Raman self-frequency shift.}
    \label{fig:soliton_spectra}
\end{figure}

Figure~\ref{fig:soliton_spectra} compares the Raman-free and Raman-enabled soliton spectra. In our simulations, the Raman-free spectrum is centered at the pump, whereas the Raman-enabled spectrum exhibits a spectral-center shift toward lower frequencies. These configurations therefore represent the spectrally unshifted and Raman-dominated shifted states, respectively.

To estimate incremental contributions, we use pump phase noise with a Lorentzian linewidth of $\Delta\nu = 2$~kHz as the reference perturbation. The added-noise simulations use shot noise, pump RIN with $S_{\mathrm{RIN}}=-140$~dBc/Hz over a 5~MHz bandwidth, EDFA ASE with a noise figure (NF) of 3~dB, and TRN. Each additional noise source is added individually to the pump-phase-noise baseline. For each noise configuration, we average over 10 independent runs with different random seeds to reduce realization-to-realization scatter in the phase-variance and PSD estimates.

Figure~\ref{fig:pipeline} summarizes how the physical noise sources are introduced and analyzed. Pump phase noise and RIN are applied to the launched pump field, EDFA ASE is added along the amplification path, and shot noise enters through the resonator loss channels. After propagation, the comb-line phase traces are unwrapped and detrended to form a phase noise matrix. Covariance analysis and eigendecomposition then identify the dominant phase-noise components and their mode profiles across the comb lines.

\subsection{Without a soliton spectral shift}

We first consider the spectrally unshifted reference state, obtained in the present second-order-dispersion model by switching off the Raman term ($f_R = 0$). Pump phase noise is initially included as the sole perturbation. We select the pump mode and 10 comb lines on either side, corresponding to $n=-10,\ldots,10$, giving 21 phase traces in total, and use them to construct the phase noise covariance matrix.

\begin{figure}
    \centering
    \includegraphics[width=\linewidth]{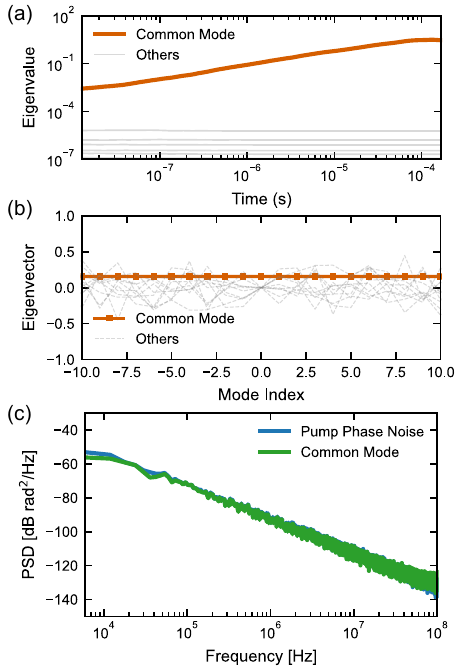}
    \caption{Subspace tracking under pump phase noise without Raman. (a) Eigenvalues versus observation time. (b) Eigenvectors across comb lines. (c) PSDs of the injected pump phase noise and recovered common-mode component.}
    \label{fig:combined_PNwoRaman}
\end{figure}

Figure~\ref{fig:combined_PNwoRaman}(a) shows one eigenvalue that grows with the observation window. The associated eigenvector in Fig.~\ref{fig:combined_PNwoRaman}(b) is flat across the comb lines, showing that the dominant component couples to each line with equal weight and therefore represents common-mode noise. In Fig.~\ref{fig:combined_PNwoRaman}(a), ``other modes'' denote the remaining eigenmodes of the covariance matrix. Their eigenvalues do not grow with the observation window, and the associated eigenvectors do not form reproducible profiles across the comb lines. These modes therefore represent the noise floor rather than resolved phase noise components.

Figure~\ref{fig:combined_PNwoRaman} (c) supports this interpretation. The phase noise component identified by subspace tracking has a PSD that matches the pump phase noise PSD. This agreement shows that pump phase noise is transferred directly to the soliton microcomb as common-mode noise, with each comb line inheriting the pump phase fluctuations.

Beyond pump phase noise, two representative additive noise sources are commonly encountered in practical microcomb systems. Shot noise is fundamentally unavoidable: vacuum fluctuations associated with optical loss enter the resonator through its dissipative channels, including external coupling and intrinsic loss, and therefore remain present even in an otherwise ideal system. EDFA ASE, by contrast, arises from a common experimental constraint. The output of a narrow-linewidth CW laser is often insufficient---after coupling and insertion losses---to sustain the intracavity power required for soliton formation. Optical amplification before the chip is therefore commonly used, and any erbium-doped fiber amplifier inevitably adds broadband spontaneous emission to the pump field in the process.

To determine whether these two additive noise sources preserve the two-component picture or introduce additional phase-noise structure, we examine shot noise and EDFA ASE separately and then in combination, with pump phase noise retained as the baseline in all cases. For each case, subspace tracking is applied independently to the simulated multi-line phase fluctuations.

\begin{figure*}
    \centering
    \includegraphics{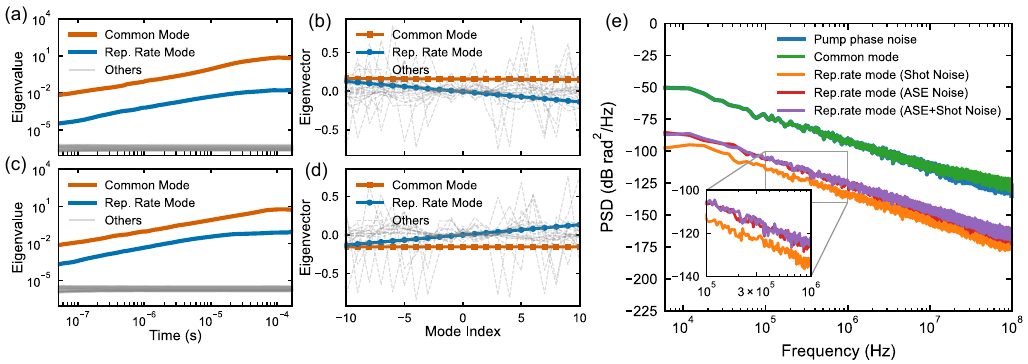}
    \caption{Subspace tracking with shot noise and EDFA ASE. (a,c) Eigenvalues versus observation time. (b,d) Eigenvectors across comb lines. (e) Phase-noise PSDs for individual and combined noise configurations.}
    \label{fig:shot_ase_combined}
\end{figure*}

When shot noise is added to the baseline with pump phase noise, the eigendecomposition changes markedly. Figure~\ref{fig:shot_ase_combined}(a) shows that a second eigenvalue grows with observation time and separates clearly from the cluster of remaining eigenvalues. The eigenvectors in Fig.~\ref{fig:shot_ase_combined}(b) identify two components: one is flat across the comb line index, retaining the common-mode profile, whereas the other varies linearly with index and crosses zero at the pump mode, a signature of a repetition-rate component. When EDFA ASE is added instead, the same qualitative change is observed. Figure~\ref{fig:shot_ase_combined}(c) shows the emergence of a second growing eigenvalue, and the corresponding eigenvectors in Fig.~\ref{fig:shot_ase_combined}(d) again exhibit a flat common-mode profile and a linear profile centered on the pump mode, characteristic of repetition-rate noise.

These results show that the configurations with shot noise and with EDFA ASE are each well described by two dominant subspace components: a common-mode component and a repetition-rate component. We project the simulated phase traces onto the corresponding eigenvectors to recover their phase noise spectra. Figure~\ref{fig:shot_ase_combined}(e) compares the pump phase noise PSD, the common-mode PSDs recovered from the two individual configurations, and the repetition-rate PSDs obtained with shot noise, EDFA ASE, and both sources together. The common-mode PSDs overlap the pump phase noise PSD, confirming that the common mode remains dominated by the pump. By contrast, the repetition-rate PSD produced by EDFA ASE exceeds that produced by shot noise across the analyzed offset frequency band.

When shot noise and EDFA ASE are added together to the baseline with pump phase noise, an independent subspace tracking analysis yields the same two component structure as in the individual configurations. The two growing eigenvalues correspond to flat common-mode and linear repetition-rate eigenvectors. This behavior reflects the resonator's dynamical mapping: both additive sources couple into the repetition-rate channel, and subspace tracking separates the resulting common-mode and repetition-rate components from the comb line phase traces without distinguishing the underlying physical sources. The repetition-rate PSD obtained with both additive sources nearly overlaps the result for the configuration with EDFA ASE in Fig.~\ref{fig:shot_ase_combined}(e), indicating that EDFA ASE, rather than shot noise, sets the repetition-rate noise floor in this comparison.

We also examine the effect of pump RIN. In a higher-RIN stress test, where $S_{\mathrm{RIN}}$ is raised from the main value of $-140$~dBc/Hz to $-110$~dBc/Hz, the eigendecomposition remains essentially identical to the case with pump phase noise only, with no discernible repetition-rate component. Thus, no RIN-induced repetition-rate contribution is resolved within the stable-soliton regime explored here. Further increasing the RIN level destabilizes the single-soliton state, rendering noise analysis inapplicable. The supporting data are given in the Supplementary Material.

Adding TRN at the level estimated for the present resonator geometry (Sec.~\ref{sec:simulation-parameters}) leaves a single growing eigenvalue with a flat eigenvector across the comb lines. The recovered common-mode PSD overlaps the pump-phase-noise-only baseline, and no repetition-rate component is detected. The corresponding eigenvalue, eigenvector, and PSD data are given in the Supplementary Material.

This mapping for the spectrally unshifted state provides the reference for the shifted state considered below: pump phase noise maps onto the common-mode component, whereas shot noise and EDFA ASE map onto the repetition-rate component. For the established single-soliton states analyzed here, adding pump RIN or TRN does not produce a resolvable change in the pump-dominated common-mode PSD or a detectable repetition-rate component.

\subsection{With a soliton spectral shift: the Raman-dominated case}

A soliton spectral shift can arise from the Raman-induced soliton self-frequency shift or from dispersive-wave recoil associated with higher-order dispersion and mode interactions \cite{yi2017single,lei2022optical}. Although these effects have different physical origins, both displace the soliton spectral center from the pump and thereby break the pump-centered spectral symmetry. Here we focus on a Raman-dominated spectrally shifted state and examine how its phase-noise structure differs from that of the unshifted reference.

\subsubsection{Noise structure of the spectrally shifted state}

\begin{figure}
    \centering
    \includegraphics[width=\linewidth]{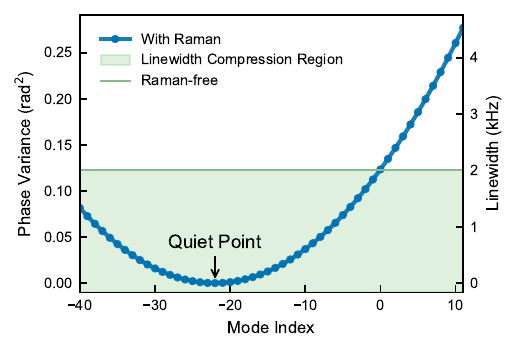}
    \caption{Comb-line phase-noise variance (left axis) and equivalent Lorentzian linewidth (right axis) with Raman self-frequency shift under pump phase noise alone. The horizontal line shows the uniform comb-line linewidth without Raman (2~kHz). The shaded region indicates linewidth compression relative to the Raman-free case.}
    \label{fig:Variance_Linewidth}
\end{figure}

When the Raman effect is included ($f_R = 0.2$), pump phase noise no longer produces a uniform linewidth distribution across the comb. Figure~\ref{fig:Variance_Linewidth} shows that the equivalent Lorentzian linewidth becomes mode dependent, with pronounced noise suppression on the red-shifted side.

The pump mode ($n=0$) retains its 2~kHz Lorentzian linewidth. Moving toward negative comb-line indices, the equivalent linewidth first decreases, reaches its minimum at $n=-22$, and then increases, forming an approximately parabolic profile. The minimum lies below the 2~kHz pump linewidth.

For the covariance analysis, we select the 37 comb lines $n=-40,\ldots,-4$, centered at $n_0=-22$. The recovered common-mode phase is referenced to this window center.

\begin{figure}
    \centering
    \includegraphics[width=\linewidth]{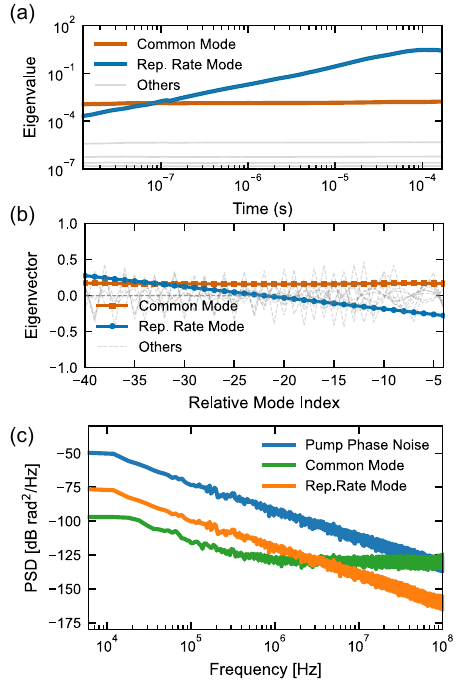}
    \caption{Subspace tracking with the Raman self-frequency shift. (a) Eigenvalues versus observation time. (b) Eigenvectors across comb lines. (c) PSDs of the recovered common-mode and repetition-rate components and the injected pump phase noise.}
    \label{fig:Raman_eigen_analysis}
\end{figure}

The eigendecomposition of the phase-noise covariance matrix is presented in Fig.~\ref{fig:Raman_eigen_analysis}(a,b). In contrast to the Raman-free case, the dominant eigenvector now varies linearly with the comb-line index, and its eigenvalue grows with the observation time. This identifies repetition-rate noise as the dominant growing phase-diffusion component. Although the second eigenvalue appears nearly flat in the figure, numerical inspection shows that it remains constant from $10^{-7}$ to $10^{-5}\,\mathrm{s}$, followed by a slight increase beyond $10^{-5}\,\mathrm{s}$. The corresponding second eigenvector has a common-mode profile, indicating an equal projection across all comb lines.

Figure~\ref{fig:Raman_eigen_analysis}(c) shows the separation of these two components: the repetition-rate component exhibits the dominant phase diffusion, whereas the common-mode component approaches a white-noise floor at high frequencies but retains a weak low-frequency contribution. This residual contribution is consistent with the slight increase of the second eigenvalue observed at longer time scales in Fig.~\ref{fig:Raman_eigen_analysis}(a). Thus, under pump phase noise alone, the Raman-shifted state contains a single resolved growing phase-diffusion degree of freedom.

\begin{figure*}
    \centering
    \includegraphics[width=\textwidth]{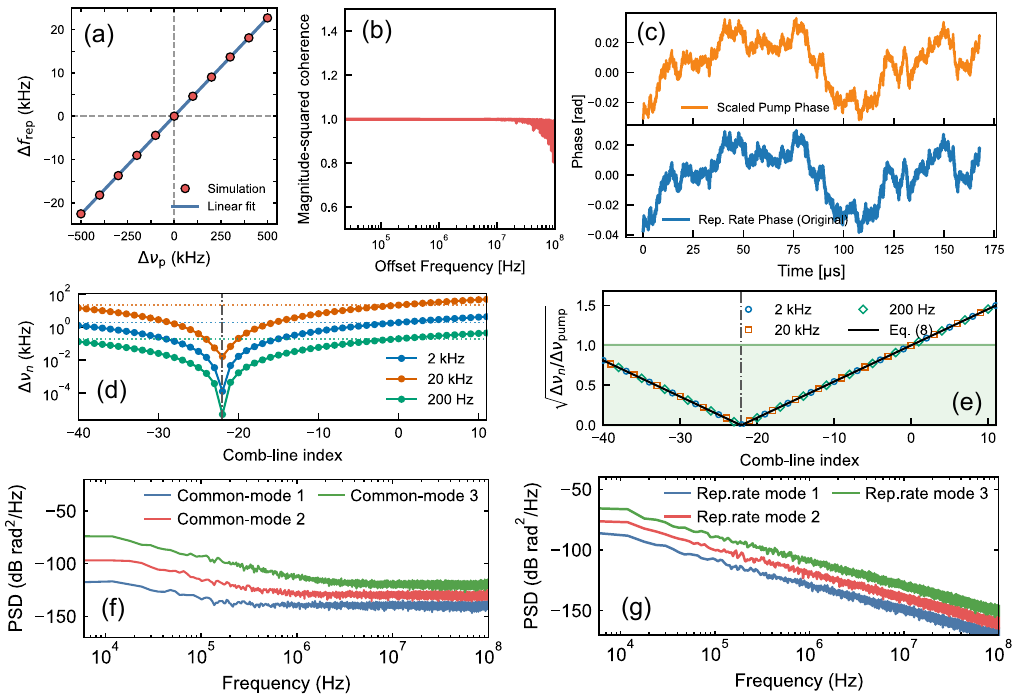}
    \caption{Raman-mediated pump-to-repetition-rate coupling.
(a) Repetition-rate shift versus pump-frequency offset and linear fit, giving $\chi=0.0453$.
(b) Magnitude-squared coherence between the pump and recovered repetition-rate phases.
(c) Scaled pump phase and recovered repetition-rate phase in time.
(d) Comb-line linewidth on a logarithmic axis; dotted horizontal lines mark the three pump linewidths and the dash-dotted vertical line marks $n_q$.
(e) The same data normalized by the respective pump linewidth; the solid black line shows the prediction of Eq.~\eqref{eq:linewidth_law}.
(f) Common-mode phase-noise PSDs re-referenced to the adopted continuous fixed point. (g) Recovered repetition-rate phase-noise PSDs.
}
    \label{fig:detuning_frep}
\end{figure*}

To identify the physical origin of the recovered repetition-rate component, we next examine how the repetition rate responds to pump-frequency perturbations. Pump phase noise can be viewed as fluctuations of the instantaneous pump frequency. With the cavity resonance frequency held fixed, a pump-frequency shift changes the pump--cavity detuning and thereby perturbs the soliton repetition rate. Figure~\ref{fig:detuning_frep}(a) shows a linear repetition-rate response to the pump-frequency offset. For the resonator parameters and stable-soliton operating point used here, a linear fit gives $\chi \equiv d\nu_{\mathrm{rep}}/d\nu_{\mathrm{pump}} = 0.0453$.

To determine whether the fluctuating repetition-rate response dynamically follows the pump noise, we compute the magnitude-squared coherence $C_{xy}(f)=|S_{xy}(f)|^2/[S_{xx}(f)S_{yy}(f)]$ between the imposed pump phase trace $x$ and the recovered repetition-rate phase trace $y$, where $S_{xy}$ is the cross-spectral density and $S_{xx}$ and $S_{yy}$ are the corresponding auto-spectral densities. A value of $C_{xy}=1$ corresponds to perfect linear coherence at that offset frequency. As shown in Fig.~\ref{fig:detuning_frep}(b), $C_{xy}(f)$ remains close to unity over almost the entire analyzed band, with only a slight reduction at the highest offset frequencies. Figure~\ref{fig:detuning_frep}(c) further shows that scaling the pump phase trace by $\chi$ closely reproduces the recovered repetition-rate phase trace in the time domain. Together, the near-unity coherence and time-domain scaling show that the single growing component is coherently driven by the pump. With $\chi$ independently obtained from the static response in Fig.~\ref{fig:detuning_frep}(a), this gives the dominant linear response, $\delta\nu_{\mathrm{rep}}=\chi\,\delta\nu_{\mathrm{pump}}$. Applying the pump-referenced comb-line relation of the elastic-tape model, $\nu_n=\nu_{\mathrm{pump}}+n\nu_{\mathrm{rep}}$, used in Ref.~\cite{lei2022optical} then gives
\begin{equation}
\label{eq:comb_response}
\delta \nu_n = \left( 1 + n\chi \right) \delta \nu_{\mathrm{pump}}
            = \chi\,(n - n_{q})\,\delta \nu_{\mathrm{pump}}.
\end{equation}

Here $n_q=-1/\chi=-22.075$ is the continuous fixed point at which the direct pump response and the Raman-mediated repetition-rate response cancel. Because the continuous fixed point lies between integer-indexed comb lines, the nearest physical line retains a finite pump-derived linewidth. The line-frequency excursion therefore scales with $|n-n_q|$, giving the parabolic pump-induced linewidth profile quantified below.

The transfer factor in Eq.~\eqref{eq:comb_response} depends only on $\chi$ and the line index, so it sets the shape of the linewidth profile but not its scale. For the Lorentzian phase-diffusion model used here, the comb-line linewidth follows
\begin{equation}
\label{eq:linewidth_law}
\Delta\nu_n=|1+n\chi|^2\,\Delta\nu_{\mathrm{pump}}=\chi^2(n-n_q)^2\,\Delta\nu_{\mathrm{pump}}.
\end{equation}
Changing the pump linewidth should therefore rescale the whole profile without moving its minimum. We tested this by repeating the simulation with pump linewidths of 200~Hz, 2~kHz, and 20~kHz, with all other parameters unchanged.

Figure~\ref{fig:detuning_frep}(d) shows the resulting profiles on a logarithmic linewidth axis, as required by the two-decade span of the pump linewidths. All three profiles reach their minimum at $n=-22$, while their absolute linewidth levels scale with the pump linewidth. Figure~\ref{fig:detuning_frep}(e) shows that normalizing each profile by its respective pump linewidth collapses the three data sets onto a single curve. Taking the square root gives $\sqrt{\Delta\nu_n/\Delta\nu_{\mathrm{pump}}}=|\chi||n-n_q|$, producing the V-shaped dependence. The solid black line shows the prediction of Eq.~\eqref{eq:linewidth_law} evaluated with $\chi=0.0453$ from the static response in Fig.~\ref{fig:detuning_frep}(a).

Applying the same subspace decomposition to the three cases separates this scaling into its common-mode and repetition-rate components. Figure~\ref{fig:detuning_frep}(g) shows that the repetition-rate PSDs retain the same spectral shape and are separated by 10~dB between successive pump-linewidth settings. This 10~dB spacing matches the decade-by-decade scaling of the injected pump phase-noise PSD, directly identifying pump phase diffusion as the source of the recovered repetition-rate component.

In Fig.~\ref{fig:detuning_frep}(f), the common-mode phase has been re-referenced to the adopted continuous fixed point, but a low-frequency excess remains. Writing $\phi_{\mathrm{rep}}=\chi\phi_{\mathrm{p}}+\epsilon$, where $\epsilon$ denotes the small deviation from the static linear response, gives $\phi_{\mathrm{cm}}=(1+n_q\chi)\phi_{\mathrm{p}}+n_q\epsilon$. The direct pump contribution and its coherent repetition-rate response cancel, whereas the residual receives a PSD weighting of approximately 487, the square of $n_q$. Thus, a small colored residual becomes visible despite a mean magnitude-squared coherence exceeding 0.999 over 10~kHz--1~MHz across all three pump linewidths. Small numerical errors may contribute to this residual. Finite-record effects may additionally limit the precision of the estimated PSD at the lowest offset frequencies.

\subsubsection{Common-mode and repetition-rate noise with additional noise sources}

\begin{figure}
    \centering
    \includegraphics[width=\linewidth]{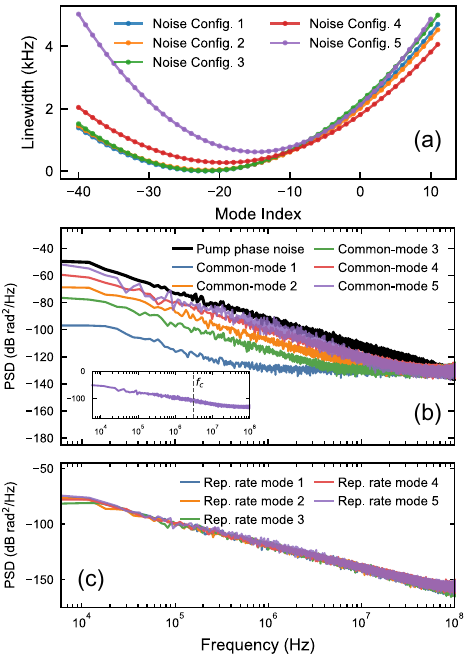}
    \caption{Phase-noise components with Raman and individually added noise sources. (a) Equivalent Lorentzian linewidth obtained by fitting the $1/f^2$ region of the phase-noise PSD from $10^5$ to $10^6$~Hz. The inset in (b) shows the TRN-dominated common-mode phase-noise PSD, with the dashed line marking the thermal corner $f_c=(2\pi\tau_d)^{-1}=3.07$~MHz. (b) Common-mode and (c) repetition-rate phase-noise PSDs. All cases include pump phase noise and Raman: 1, baseline; 2, shot noise; 3, RIN; 4, EDFA ASE; 5, TRN.}
    \label{fig:VarianceandNoise}
\end{figure}

We next add shot noise, EDFA ASE, RIN, or TRN individually to the case with Raman and pump phase noise. Figure~\ref{fig:VarianceandNoise}(a) shows the equivalent Lorentzian linewidths for the five configurations, obtained by fitting the $1/f^2$ region of the phase-noise PSD from $10^5$ to $10^6$~Hz. The linewidth profiles remain approximately parabolic, while the added noise sources raise the minimum and modify its position to different extents.

Subspace tracking first tests whether the added noise sources introduce new growing components. In the case with Raman and pump phase noise only, the common-mode eigenvalue shows only a very slight increase at long observation times, consistent with the weak low-frequency contribution and the high-frequency white-noise floor described above. When shot noise, EDFA ASE, RIN, or TRN is added, the common-mode profile becomes associated with a growing eigenvalue. No third growing residual component appears in the configurations considered here, so the comb-line phase noise remains captured by the common-mode and repetition-rate terms of Eq.~\eqref{eq:elastic_tape}.

Figures~\ref{fig:VarianceandNoise}(b) and \ref{fig:VarianceandNoise}(c) quantify these two components in the PSD domain. Despite averaging over ten independent realizations, the TRN-dominated common-mode PSD retains slightly greater finite-record scatter below the thermal corner $f_c=(2\pi\tau_d)^{-1}=3.07$~MHz, as marked in the inset of Fig.~\ref{fig:VarianceandNoise}(b). Although a new Langevin increment is applied every roundtrip, the OU detuning retains memory over $\tau_d=51.8$~ns. The enhanced scatter below $f_c$ is therefore consistent with each finite record containing fewer effectively independent thermal fluctuations than its sample count suggests. Above $f_c$, the TRN contribution rolls off, and the residual scatter becomes comparable to that of the other noise configurations. The fitted common-mode Lorentzian linewidths are 0.26~Hz for the baseline and approximately 56, 9.3, 299, and 672~Hz for the configurations with added shot noise, RIN, EDFA ASE, and TRN, respectively. EDFA ASE and TRN produce the largest common-mode linewidths in this comparison, both in the hundreds-of-hertz range. This ranking can change with the injected noise strengths. Figure~\ref{fig:VarianceandNoise}(c) shows that the repetition-rate PSDs nearly overlap across all five configurations. Raman-mediated transfer of pump phase noise therefore dominates the repetition-rate component, while the added sources make only small contributions to its PSD.

The different effects of RIN and TRN in the unshifted and Raman-shifted soliton states reflect how the soliton spectral center responds to noise. In the Raman-shifted state, pump RIN modulates the intracavity soliton intensity and pulse shape, thereby affecting the Raman self-frequency shift. TRN shifts the cavity resonance and thus changes the pump--cavity detuning, accessing the same detuning-to-repetition-rate pathway as pump-frequency noise. In either case, dispersion converts the resulting Raman-shift fluctuations into group-velocity and hence repetition-rate fluctuations. Under the conditions studied here, these additional repetition-rate contributions are small, particularly for RIN, and pump phase noise remains dominant. In the unshifted soliton state, this conversion pathway is absent, consistent with our observation that neither RIN nor TRN produces a resolved repetition-rate noise component.

\section{Conclusion}

By combining subspace tracking with multi-source Ikeda-map simulations, we resolved the common-mode and repetition-rate phase-noise components of soliton microcombs and examined their physical origins. Switching individual noise sources on and off, and comparing the spectrally symmetric soliton with the spectrally shifted state realized here through the Raman response, revealed which components were driven by each perturbation. Across all cases, no additional growing component was observed.

For the spectrally symmetric soliton, pump phase noise appeared entirely as common-mode noise, whereas shot noise and EDFA ASE drove repetition-rate noise, with EDFA ASE contributing more strongly in this comparison. In the same state, adding pump RIN or TRN did not produce a resolvable change in the pump-dominated common-mode PSD or a detectable repetition-rate component. In the Raman-dominated shifted state analyzed here, coherent pump-to-repetition-rate conversion emerged, collapsing the common-mode contribution to a floor far below the pump phase noise. The superposition of the direct and Raman-mediated responses produced an approximately parabolic linewidth dependence on comb-line index, with the nearest comb line to the fixed point reaching a linewidth below that of the pump. Raman-mediated pump-to-repetition-rate conversion remained dominant across the added-source configurations. These sources made only small additional contributions to the repetition-rate PSD while raising the common-mode floor, most strongly for TRN and EDFA ASE.

More broadly, these results show that low-noise microcomb design must address both the physical noise sources and the intracavity dynamics that redistribute their fluctuations across the comb. The source-to-component assignments identified in this study and the shift-enabled transfer of pump phase noise from the common mode to the repetition rate are mechanism-level conclusions. The coupling coefficient and quiet-point position, by contrast, depend on the material response and soliton operating point. The absolute PSD and linewidth levels additionally depend on the applied source levels. The combined framework therefore provides a route to identify the dominant coherence-limiting mechanism and guide low-noise soliton-microcomb design. The present treatment of thermal noise is limited to first-order cavity-detuning fluctuations. Extending the analysis to additional thermal contributions and comb-line intensity noise is a natural direction for the source-resolved framework developed here.

Experimental validation requires mapping the selected comb lines to distinguishable detection channels and acquiring their phase trajectories synchronously over a common observation interval. One possible implementation is parallel optical demultiplexing followed by coherent down-conversion and synchronous sampling of the selected lines. Alternatively, dual-comb multiheterodyne detection maps the paired comb lines directly onto distinct radio-frequency beat notes. By contrast, sequential tuning of a single optical filter followed by delayed self-heterodyne measurement can characterize individual-line linewidths or phase-noise spectra but cannot provide the synchronous multi-line phase traces required to construct the inter-line covariance matrix. Because the reference comb enters every recovered trace, its own common-mode and repetition-rate noise must be sufficiently low or independently characterized. The physical mode-index mapping between the paired comb lines must also be known.

\section{Numerical implementation}
\label{sec:numerical-implementation}

\subsection{Simulation parameters and soliton-state preparation}
\label{sec:simulation-parameters}

The single-soliton initial states are generated using the continuous-wave triggering method of Ref.~\cite{wu2025deterministic}. The resonator parameter set used for soliton generation is representative of an integrated Si$_3$N$_4$ microring~\cite{twayana2023advanced}, with a pump wavelength of 1550~nm, $\mathrm{FSR}=100$~GHz, $n_g=2.1$, $\beta_2=-150$~ps$^2$/km, $\gamma=1.1$~W$^{-1}$m$^{-1}$, $\theta=3.4\times10^{-4}$, and $\alpha=9$~dB/m (equivalent loaded $Q\simeq3.7\times10^6$).

For the TRN calculation, we use $T=300$~K, $\rho=3.29\times10^{3}$~kg\,m$^{-3}$, $C_p=800$~J\,kg$^{-1}$\,K$^{-1}$, $\kappa=30$~W\,m$^{-1}$\,K$^{-1}$, and $dn_{\mathrm{eff}}/dT=2.45\times10^{-5}$~K$^{-1}$ for Si$_3$N$_4$\cite{huang2019thermorefractive}. The effective optical-mode volume inferred from the simulated resonator is $V_{\mathrm{eff}}=1.274\times10^{-15}$~m$^3$. For the radial thermal-diffusion scale, we adopt $d_r=0.8~\mu$m from Ref.~\cite{zhou2026overcomingTRN}, yielding $\sigma_\nu=43.4$~kHz and $\tau_d=51.8$~ns.

The Raman-free and Raman-enabled soliton states are generated separately. Their launched pump powers are both 100~mW, and their final pump--cavity frequency detunings are $-200$ and $-220$~MHz, respectively. The fast-time grid contains $N=2048$ points per roundtrip. The simulations are propagated over $2^{24}$ roundtrips, with the comb-line phases recorded once every 128 roundtrips to reduce memory consumption.

All noise sources are introduced only after a stationary single-soliton state has been prepared.

The Raman-free case uses $f_R=0$. For the Raman-enabled case, $f_R=0.2$ follows the value reported for Si$_3$N$_4$ in Ref.~\cite{karpov2016raman}, and we use $\tau_1=30$~fs and $\tau_2=120$~fs as effective response times.

\subsection{Pump phase noise}

The pump phase follows a Wiener process\cite{mandel1996opticalWiener}: $d\phi(t) = \sqrt{2\pi \Delta\nu} \cdot dW(t)$, where $\Delta\nu$ is the Lorentzian linewidth and $W(t)$ is the standard Wiener process. Because the coherence time of a typical pump laser far exceeds the cavity roundtrip time, the phase noise is updated once per roundtrip. The discrete phase increment is:

\begin{equation}
\delta\phi = \sqrt{\frac{2\pi \Delta\nu}{\mathrm{FSR}}} \cdot \xi, \quad \xi \sim \mathcal{N}(0,1)
\end{equation}

The accumulated stochastic pump phase fluctuation $\phi_m = \sum_{j=1}^{m} \delta\phi_j$ evolves as a discrete random walk and enters the boundary condition through Eq.~\eqref{eq:input_field}.

\subsection{Shot noise}

Shot noise is modeled within the semiclassical framework. The continuous vacuum fluctuation field satisfies $\langle \delta E^*(t) \delta E(t + \tau) \rangle = \frac{h\nu_p}{2}\delta(\tau)$\cite{paschotta2004noiseshot}, where $h$ is the Planck constant and $\nu_p$ is the pump frequency. Within a single discrete time bin $\Delta t = 1/(N\cdot \mathrm{FSR})$, the complex noise samples are constructed as:

\begin{equation}
\delta E_k^{\mathrm{(vac)}} = \sqrt{\frac{h\nu_p N \cdot \mathrm{FSR}}{4}} (\xi_1 + i\xi_2), \quad \xi_{1,2} \sim \mathcal{N}(0,1)
\end{equation}

where $N$ is the number of discrete sampling points per roundtrip and $k = 1, 2, \dots, N$ is the fast-time grid index. To enhance numerical efficiency, the distributed vacuum fluctuations are equivalently aggregated at the boundary, weighted by the total dissipation coefficient\cite{yariv2000universal}:

\begin{equation}
\Gamma_{\mathrm{tot}} \equiv \frac{\kappa_{\mathrm{tot}}}{\mathrm{FSR}} = \frac{2\pi\nu_p}{\mathrm{FSR}}\left(\frac{1}{Q_{\mathrm{ex}}} + \frac{1}{Q_{\mathrm{in}}}\right) \approx -\ln(1-\theta) + \alpha L
\end{equation}

yielding the shot-noise field $\delta E_k^{\mathrm{(shot)}} = \sqrt{\Gamma_{\mathrm{tot}}} \cdot \delta E_k^{\mathrm{(vac)}}$ injected at the boundary in Eq.~\eqref{eq:boundary}.

\subsection{Amplified spontaneous emission}

When the pump is amplified by an EDFA with power gain $G$ and spontaneous emission factor $n_{\mathrm{sp}}$, the single-sided ASE power spectral density per polarization is $S_{\mathrm{ASE}} = n_{\mathrm{sp}} h \nu_p (G - 1)$\cite{agrawal2012fiberEDFA}. For a specified noise figure, we use the high-gain relation $n_{\mathrm{sp}}\simeq F_{\mathrm{EDFA}}/2$, where $F_{\mathrm{EDFA}}=10^{\mathrm{NF}/10}$ is the linear noise figure. In the simulations, the coherent pump amplitude is fixed by the launched power $\bar{P}_{\mathrm{in}}$ that generates the stable soliton state in Eq.~\eqref{eq:input_field}. The EDFA parameters therefore determine only the additive ASE level, keeping the soliton operating point identical between ASE and non-ASE configurations. The discrete complex noise field preserving the correct noise power is:

\begin{equation}
\delta E_k^{\mathrm{(ASE)}} = \sqrt{\frac{n_{\mathrm{sp}} h \nu_p (G - 1) N \cdot \mathrm{FSR}}{2}} (\xi_1 + i\xi_2)
\end{equation}

where $\xi_1$ and $\xi_2$ are independent standard normal variables. This gives $\langle|\delta E_k^{\mathrm{(ASE)}}|^2\rangle = S_{\mathrm{ASE}}N\,\mathrm{FSR}$, and the field is additive to the launched pump field in Eq.~\eqref{eq:input_field}.

\subsection{Relative intensity noise}

RIN characterizes low-frequency power fluctuations of the pump laser, with $\eta(t) = \delta P(t)/\bar{P}_{\mathrm{in}}$ and single-sided PSD $S_{\mathrm{RIN}}(f)$. In the simulations, the RIN spectrum is taken to be flat, $S_{\mathrm{RIN}}(f)=S_{\mathrm{RIN}}$, over the target bandwidth $B_{\mathrm{target}}$. Numerically, the fluctuation is updated every $\mathrm{FSR}/(2B_{\mathrm{target}})$ roundtrips and held constant between updates, with samples drawn as:

\begin{equation}
\delta\eta_m = \sqrt{S_{\mathrm{RIN}} B_{\mathrm{target}}}\,\xi_m, \quad \xi_m \sim \mathcal{N}(0,1)
\end{equation}

The instantaneous pump power is then $P_m = \bar{P}_{\mathrm{in}}(1+\delta\eta_m)$.

\subsection{Thermorefractive noise}

Thermodynamic temperature fluctuations sampled by the optical mode shift the cavity-resonance frequency through the thermo-optic effect\cite{gorodetsky2004fluctuations,huang2019thermorefractive}. Following Xue \textit{et al.}\cite{xue2016thermaltuning}, we neglect the thermal-expansion contribution in this first-order approximation. The temperature and resonance-frequency fluctuations are then related by

\begin{equation}
\sigma_T^2 = \frac{k_B T^2}{\rho C_p V_{\mathrm{eff}}}, \qquad
\sigma_\nu = \left|\frac{d\nu_{\mathrm{res}}}{dT}\right|\sigma_T
\simeq \frac{\nu_p}{n_g}\left|\frac{dn_{\mathrm{eff}}}{dT}\right|\sigma_T .
\end{equation}

The final expression is the group-index form of the thermorefractive temperature-to-frequency conversion and is consistent with Eq.~(7) of Ref.~\cite{zhou2026overcomingTRN} under the approximation $n_g\simeq n_{\mathrm{eff}}$. Here, $V_{\mathrm{eff}}\approx A_{\mathrm{eff}}L$, where $A_{\mathrm{eff}}$ is inferred from the nonlinear waveguide parameters. We use the Si$_3$N$_4$ thermo-optic coefficient as a first-order estimate of $dn_{\mathrm{eff}}/dT$. With the thermal diffusivity $D=\kappa/(\rho C_p)$, the effective radial thermal-diffusion time is estimated as\cite{huang2019thermorefractive,zhou2026overcomingTRN}

\begin{equation}
\tau_d=\left(\frac{\pi}{4}\right)^{1/3}\frac{d_r^2}{D}
=\left(\frac{\pi}{4}\right)^{1/3}\frac{\rho C_p}{\kappa}d_r^2.
\end{equation}

Here, $d_r$ is the effective radial optical-mode dimension.

These relations determine the values of $\sigma_\nu$ and $\tau_d$ used in the simulations. Within the first-order single-pole approximation, the resonance-frequency fluctuation is modeled as a stationary Ornstein--Uhlenbeck process, i.e., as the stationary solution of the Langevin equation

\begin{equation}
d\delta\nu(t)=-\frac{\delta\nu(t)}{\tau_d}\,dt
+\sqrt{\frac{2\sigma_\nu^2}{\tau_d}}\,dW(t),
\end{equation}

where $W(t)$ is a standard Wiener process. The corresponding single-sided frequency-noise PSD is

\begin{equation}
S_{\delta\nu}(f) = \frac{4\sigma_\nu^{2}\tau_d}{1+(2\pi f\tau_d)^{2}}.
\end{equation}

The OU process is propagated using its exact discrete-time transition over one cavity roundtrip\cite{gillespie1996exact}:

\begin{equation}
\begin{aligned}
\delta\nu_{m+1} &= a\,\delta\nu_m
+ \sigma_\nu\sqrt{1-a^{2}}\,\xi_m, \\
a &= e^{-1/(\tau_d\,\mathrm{FSR})}, \\
\xi_m &\sim \mathcal{N}(0,1).
\end{aligned}
\end{equation}

The fluctuation enters the boundary condition of Eq.~\eqref{eq:boundary} as $\delta_m^{\mathrm{TRN}}=2\pi\delta\nu_m/\mathrm{FSR}$. The model thus represents TRN as a fluctuation of the cavity resonance frequency.

\subsection{Raman response function}

The normalized nonlinear response function decomposes the total nonlinearity into instantaneous and delayed contributions\cite{AGRAWAL201927Raman}:

\begin{equation}
R(t) = (1 - f_R)\delta(t) + f_R h_R(t)
\end{equation}

where the Raman temporal response function is:

\begin{equation}
h_R(t) = \frac{\tau_1^2 + \tau_2^2}{\tau_1 \tau_2^2} \exp\left(-\frac{t}{\tau_2}\right) \sin\left(\frac{t}{\tau_1}\right), \quad t \geq 0
\end{equation}

Here, $\tau_1$ and $\tau_2$ are effective response times controlling the oscillation and decay of $h_R(t)$, respectively. Consistent with Eq.~\eqref{eq:propagation}, the delayed Raman-weighted intensity is evaluated in the frequency domain as:
\begin{equation}
|E(t)|_{R}^2 = \mathcal{F}^{-1} \left\{ \mathcal{F}\left\{ h_R(t) \right\} \cdot \mathcal{F}\left\{ |E(t)|^2 \right\} \right\}.
\end{equation}

The nonlinear phase step then uses $(1-f_R)|E(t)|^2+f_R|E(t)|_{R}^2$, matching the response term in Eq.~\eqref{eq:propagation}.

\section*{Supplementary Material}

The Supplementary Material presents subspace tracking results for RIN and TRN in the spectrally symmetric (unshifted) single-soliton state and compares stable and unstable soliton propagation under different RIN levels.

\begin{acknowledgments}
This work was funded in part by the European Union under the Marie Sk{\l}odowska-Curie Actions Grant Agreement No. 101120422, Quantum Enhanced Optical Communication Network Security (QuNEST).
\end{acknowledgments}

\renewcommand{\refname}{References}
\bibliography{sample}

% Supplementary Material starts as a separate part after the main article.
\onecolumngrid
\clearpage
\begin{center}
  {\large\bfseries Supplementary Material\par}
\end{center}
\raggedbottom
\setcounter{figure}{0}
\setcounter{table}{0}
\setcounter{equation}{0}
\setcounter{section}{0}
\setcounter{subsection}{0}
\renewcommand{\thefigure}{S\arabic{figure}}
\renewcommand{\thetable}{S\arabic{table}}
\renewcommand{\theequation}{S\arabic{equation}}
\renewcommand{\thesection}{S\arabic{section}}
\newcommand{\SIfiguretextfont}{\footnotesize}
\makeatletter
\newenvironment{suppfigure}{%
  \par\addvspace{4pt}\noindent\begin{minipage}{\linewidth}%
  \def\@captype{figure}%
}{\end{minipage}\par\addvspace{4pt}}
\makeatother
\makeatletter
\@ifundefined{theHfigure}{}{\renewcommand{\theHfigure}{supp.\arabic{figure}}}
\@ifundefined{theHtable}{}{\renewcommand{\theHtable}{supp.\arabic{table}}}
\@ifundefined{theHequation}{}{\renewcommand{\theHequation}{supp.\arabic{equation}}}
\@ifundefined{theHsection}{}{\renewcommand{\theHsection}{supp.\arabic{section}}}
\makeatother

This Supplementary Material tests whether relative-intensity noise (RIN) or
thermorefractive noise (TRN) adds a resolved phase-noise component to an
established, spectrally unshifted single-soliton state. Over the levels tested
here, neither source introduces a new resolved subspace component beyond the
pump-dominated common mode, while a much stronger RIN of
$S_{\mathrm{RIN}}=-80$~dBc/Hz destroys the single-soliton state altogether.
These results concern noise transfer after a stable single soliton has formed;
how the same sources act during soliton preparation is a separate question,
addressed qualitatively in Sec.~\ref{sec:preparation}.

\section{Simulation scope and comparison protocol}

The direct subspace analysis reported below is carried out in the spectrally
unshifted reference state, in which both mechanisms that could displace the
soliton spectral center from the pump are absent: the Raman response is
switched off ($f_R=0$), and the model of the main text carries no third-order
dispersion ($\beta_3=0$), so no dispersive-wave recoil arises either.

All simulations retain pump phase noise corresponding to a linewidth of
$\Delta\nu=2$~kHz. The RIN subspace analysis uses two levels,
$S_{\mathrm{RIN}}=-140$ and $-110$~dBc/Hz, over a target bandwidth of 5~MHz. A
third level, $S_{\mathrm{RIN}}=-80$~dBc/Hz, is used to test the stability of
single-soliton propagation under stronger RIN. The TRN is specified by
$\sigma_\nu=43.4$~kHz and $\tau_d=51.8$~ns. Apart from the noise
source under study, the resonator parameters, mean pump power, pump--cavity
detuning, initial single-soliton field, sampling interval, and selected
comb-line set are kept fixed. As in Fig.~3 of the main text, the covariance
matrix is built from the pump mode and ten comb lines on either side, that is,
21 phase trajectories with $n=-10,\ldots,10$. The subspace results are obtained
from ten independent noise realizations. The propagation maps show
representative single realizations with the corresponding stable or unstable
behavior.

A phase-noise component is identified as resolved only when its eigenvalue
keeps growing with observation time and the corresponding eigenvector shows a
reproducible spatial profile across the selected comb lines. This criterion
prevents an eigenvector produced by finite sampling from being read as a noise
component when no corresponding diffusive eigenvalue is present.

\section{RIN subspace analysis in the spectrally unshifted single-soliton state}

At $S_{\mathrm{RIN}}=-140$~dBc/Hz, Fig.~\ref{fig:rin140_subspace}(a)
shows that only one eigenvalue grows with observation time, while the remaining
eigenvalues stay near the non-growing noise floor. The corresponding
eigenvector in Fig.~\ref{fig:rin140_subspace}(b) is nearly constant across the
selected comb lines and is therefore identified as the common-mode phase-noise
component; no reproducible linear eigenvector, which would signal a
repetition-rate component, is resolved. In
Fig.~\ref{fig:rin140_subspace}(c), the projected common-mode phase-noise PSD
nearly overlaps the injected pump phase-noise PSD over the analyzed
offset-frequency range.

\begin{suppfigure}
  \centering
  \includegraphics[width=0.60\linewidth]{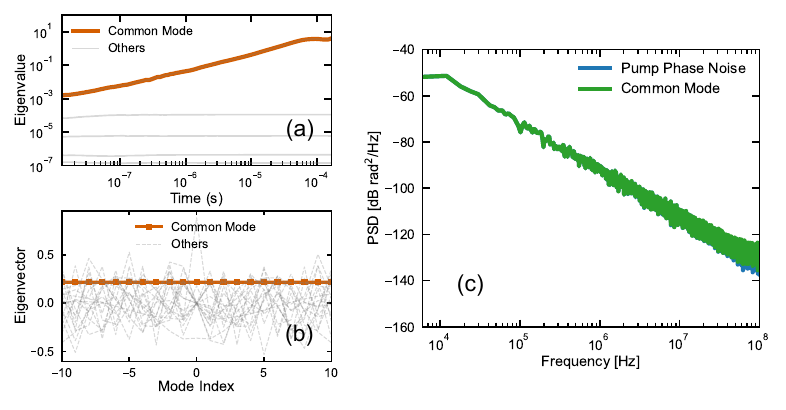}
  \caption{Subspace tracking at $S_{\mathrm{RIN}}=-140$~dBc/Hz in the
  spectrally unshifted single-soliton state. (a) Eigenvalues versus observation
  time. (b) Eigenvectors across the selected comb lines. (c) PSDs of the
  injected pump phase noise and the recovered common-mode phase-noise
  component.}
  \label{fig:rin140_subspace}
  \par\smallskip
  \begin{minipage}{\linewidth}
    \SIfiguretextfont\raggedright
    Alt text: Three line plots for relative intensity noise at -140 dBc/Hz
    show one eigenvalue growing with observation time, a flat common-mode
    eigenvector, and a recovered common-mode spectrum overlapping the injected
    pump phase-noise spectrum. No repetition-rate or residual component is
    resolved.
  \end{minipage}
\end{suppfigure}

Raising $S_{\mathrm{RIN}}$ to $-110$~dBc/Hz, 30~dB above the baseline level,
reproduces this picture unchanged: Fig.~\ref{fig:rin110_subspace} again shows a
single growing eigenvalue with a flat eigenvector, no growing eigenvalue with a
linear eigenvector, and a common-mode PSD that follows the injected pump
phase-noise PSD. Thus, within the stable single-soliton propagation range
examined here, increasing the RIN from $-140$ to $-110$~dBc/Hz does not produce
a resolved RIN-induced repetition-rate phase-noise component.

\begin{suppfigure}
  \centering
  \includegraphics[width=0.75\linewidth]{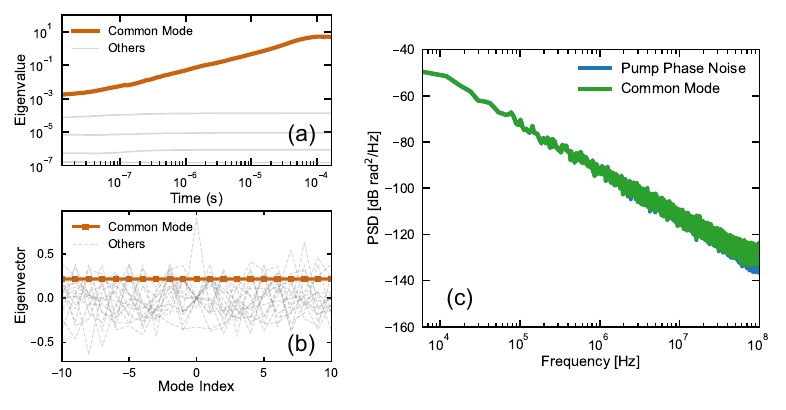}
  \caption{Subspace tracking at $S_{\mathrm{RIN}}=-110$~dBc/Hz in the
  spectrally unshifted single-soliton state. (a) Eigenvalues versus observation
  time. (b) Eigenvectors across the selected comb lines. (c) PSDs of the
  injected pump phase noise and the recovered common-mode phase-noise
  component.}
  \label{fig:rin110_subspace}
  \par\smallskip
  \begin{minipage}{\linewidth}
    \SIfiguretextfont\raggedright
    Alt text: Three line plots for relative intensity noise at -110 dBc/Hz
    show one eigenvalue growing with observation time, a flat common-mode
    eigenvector, and a recovered common-mode spectrum overlapping the injected
    pump phase-noise spectrum. No repetition-rate or residual component is
    resolved.
  \end{minipage}
\end{suppfigure}

\section{TRN subspace analysis in the spectrally unshifted single-soliton state}

Figure~\ref{fig:trn_subspace} shows the direct subspace analysis with
thermorefractive noise added. Only one eigenvalue grows with observation time,
its eigenvector stays flat across the selected comb lines, and the recovered
common-mode phase-noise PSD nearly overlaps the pump-phase-noise-only baseline
over the analyzed offset-frequency range; the remaining eigenvalues stay
non-growing. No additional repetition-rate component and no other residual
phase-noise component are resolved.

For a spectrally unshifted and already established single-soliton state,
therefore, the present level of thermorefractive detuning perturbation does not
change the common-mode subspace structure set by the pump phase noise. This does
not exclude a measurable effect of TRN at other operating points, over lower
offset-frequency ranges, or under a different definition of the effective
linewidth.

\begin{suppfigure}
  \centering
  \includegraphics[width=0.60\linewidth]{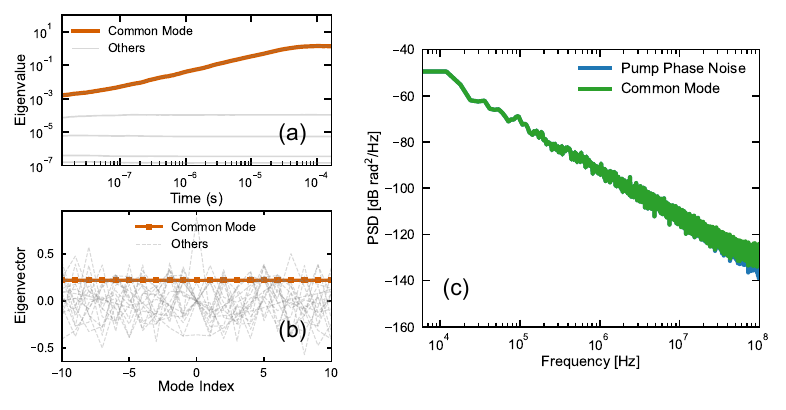}
  \caption{Subspace tracking with thermorefractive noise in the
  spectrally unshifted single-soliton state. (a) Eigenvalues versus observation
  time. (b) Eigenvectors across the selected comb lines. (c) PSDs of the
  injected pump phase noise, the recovered common-mode phase-noise component,
  and the pump-phase-noise-only baseline.}
  \label{fig:trn_subspace}
  \par\smallskip
  \begin{minipage}{\linewidth}
    \SIfiguretextfont\raggedright
    Alt text: Three line plots with thermorefractive noise show one growing
    eigenvalue, a flat common-mode eigenvector, and close overlap between the
    recovered common-mode and injected pump phase-noise spectra. No additional
    repetition-rate or residual component is resolved.
  \end{minipage}
\end{suppfigure}

\section{Representative stable and unstable propagation under pump RIN}

The subspace analysis above presumes that the underlying single-soliton state
survives the applied RIN. Figure~\ref{fig:rin_propagation} checks this premise
directly on representative propagation runs.

At $S_{\mathrm{RIN}}=-110$~dBc/Hz, the higher of the two levels used in the
subspace analysis, the localized single soliton in
Fig.~\ref{fig:rin_propagation}(a) remains stable over the $2^{20}$ roundtrips
shown, consistent with the absence of any resolved repetition-rate component in
Fig.~\ref{fig:rin110_subspace}. When the RIN is increased further to
$S_{\mathrm{RIN}}=-80$~dBc/Hz, Fig.~\ref{fig:rin_propagation}(b) shows that the
localized structure is lost, so the premise of the phase-noise analysis no
longer holds. These representative runs delimit the scope of the analysis; they
do not determine a critical RIN threshold.

\begin{suppfigure}
  \centering
  \includegraphics[width=0.60\linewidth]{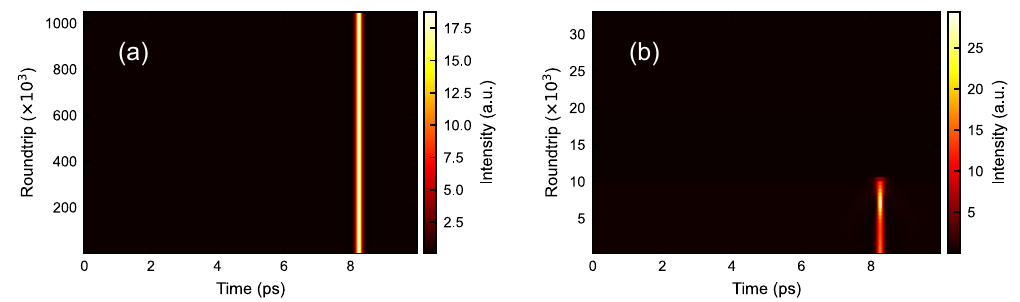}
  \caption{Representative evolution of the intracavity intensity
  $|E_m(t)|^2$ under pump RIN. (a) Stable single-soliton propagation at
  $S_{\mathrm{RIN}}=-110$~dBc/Hz over $2^{20}$ roundtrips. (b) Loss of the
  single-soliton state at $S_{\mathrm{RIN}}=-80$~dBc/Hz, occurring within
  $2^{15}$ roundtrips.}
  \label{fig:rin_propagation}
  \par\smallskip
  \begin{minipage}{\linewidth}
    \SIfiguretextfont\raggedright
    Alt text: Two intensity heat maps compare soliton propagation under two
    relative-intensity-noise levels. At -110 dBc/Hz, a localized pulse
    persists throughout the simulation; at -80 dBc/Hz, the localized structure
    disappears after about ten thousand roundtrips, indicating loss of the
    single-soliton state.
  \end{minipage}
\end{suppfigure}

\section{Boundary between soliton preparation and post-formation noise transfer}
\label{sec:preparation}

The present work addresses phase-noise transfer in a single soliton that has
already been established and is held at a fixed mean operating point. Within
that scope, adding RIN or TRN at the levels used here leaves the subspace
decomposition unchanged
(Figs.~\ref{fig:rin140_subspace}--\ref{fig:trn_subspace}). The result does not
extend to how that state is reached. In a complementary test we ran the
conventional detuning scan with the same sources switched on, and scan settings
that reach a single soliton without the added source could instead settle into
a multi-soliton state once RIN or TRN was present. Both sources therefore
influence the outcome of soliton preparation. That regime lies outside the
scope of the present analysis and is not pursued further here.

\end{document}